\documentclass{article}
\usepackage[mathscr]{eucal}
\usepackage[cmex10]{amsmath}
\usepackage{epsfig,epsf,psfrag}
\usepackage{amssymb,amsmath,amsthm,amsfonts,latexsym}
\usepackage{amsmath,graphicx,bm,xcolor,url,overpic}
\usepackage{fixltx2e}
\usepackage{array}
\usepackage{verbatim}
\usepackage{bm}
\usepackage{algorithmic}
\usepackage{algorithm}
\usepackage{verbatim}
\usepackage{textcomp}
\usepackage{mathrsfs}
\usepackage{epstopdf}

\newcommand{\openone}{\leavevmode\hbox{\small1\normalsize\kern-.33em1}}

\catcode`~=11 \def\UrlSpecials{\do\~{\kern -.15em\lower .7ex\hbox{~}\kern .04em}} \catcode`~=13 

\allowdisplaybreaks[4]

\newcommand{\nn}{\nonumber}

\newcommand{\calA}{\mathcal{A}}

\newcommand{\calF}{\mathcal{F}}

\newcommand{\calP}{\mathcal{P}}

\newcommand{\calS}{\mathcal{S}}

\newcommand{\calV}{\mathcal{V}}

\newcommand{\calX}{\mathcal{X}}
\newcommand{\calY}{\mathcal{Y}}

\newcommand{\ba}{\mathbf{a}}

\newcommand{\bs}{\mathbf{s}}
\newcommand{\bS}{\mathbf{S}}

\newcommand{\bv}{\mathbf{v}}
\newcommand{\bV}{\mathbf{V}}

\newcommand{\rmd}{\mathrm{d}}

\newcommand{\rme}{\mathrm{e}}

\newcommand{\rml}{\mathrm{l}}

\newcommand{\rmP}{\mathrm{P}}

\newcommand{\rmT}{\mathrm{T}}

\newcommand{\rmV}{\mathrm{V}}

\newcommand{\bbN}{\mathbb{N}}

\newcommand{\bbR}{\mathbb{R}}

\DeclareMathAlphabet{\mathbsf}{OT1}{cmss}{bx}{n}
\DeclareMathAlphabet{\mathssf}{OT1}{cmss}{m}{sl}

\DeclareSymbolFont{bsfletters}{OT1}{cmss}{bx}{n}  
\DeclareSymbolFont{ssfletters}{OT1}{cmss}{m}{n}
\DeclareMathSymbol{\bsfGamma}{0}{bsfletters}{'000}
\DeclareMathSymbol{\ssfGamma}{0}{ssfletters}{'000}
\DeclareMathSymbol{\bsfDelta}{0}{bsfletters}{'001}
\DeclareMathSymbol{\ssfDelta}{0}{ssfletters}{'001}
\DeclareMathSymbol{\bsfTheta}{0}{bsfletters}{'002}
\DeclareMathSymbol{\ssfTheta}{0}{ssfletters}{'002}
\DeclareMathSymbol{\bsfLambda}{0}{bsfletters}{'003}
\DeclareMathSymbol{\ssfLambda}{0}{ssfletters}{'003}
\DeclareMathSymbol{\bsfXi}{0}{bsfletters}{'004}
\DeclareMathSymbol{\ssfXi}{0}{ssfletters}{'004}
\DeclareMathSymbol{\bsfPi}{0}{bsfletters}{'005}
\DeclareMathSymbol{\ssfPi}{0}{ssfletters}{'005}
\DeclareMathSymbol{\bsfSigma}{0}{bsfletters}{'006}
\DeclareMathSymbol{\ssfSigma}{0}{ssfletters}{'006}
\DeclareMathSymbol{\bsfUpsilon}{0}{bsfletters}{'007}
\DeclareMathSymbol{\ssfUpsilon}{0}{ssfletters}{'007}
\DeclareMathSymbol{\bsfPhi}{0}{bsfletters}{'010}
\DeclareMathSymbol{\ssfPhi}{0}{ssfletters}{'010}
\DeclareMathSymbol{\bsfPsi}{0}{bsfletters}{'011}
\DeclareMathSymbol{\ssfPsi}{0}{ssfletters}{'011}
\DeclareMathSymbol{\bsfOmega}{0}{bsfletters}{'012}
\DeclareMathSymbol{\ssfOmega}{0}{ssfletters}{'012}

\newcommand{\hatS}{\hat{S}}

\newcommand{\hatV}{\hat{V}}

\newtheorem{theorem}{Theorem}

\newtheorem{definition}{Definition}

\usepackage{spconf,amsmath,graphicx}
\usepackage[utf8]{inputenc} 
\usepackage[T1]{fontenc}    
\usepackage{url,bbm}            
\usepackage{booktabs}       
\usepackage{amsfonts,algorithm,algorithmic}       
\usepackage{nicefrac}       
\usepackage{microtype}      

\usepackage{cite}
\usepackage{subfigure,transparent,color,graphicx} 

\allowdisplaybreaks[2]
\newcommand{\bbo}{\mathbbm{1}}

\title{Resolution Limits of 20 Questions Search Strategies for Moving Targets}
\twoauthors
 {Lin Zhou \sthanks{Supported by NKRDC under Grant 2020YFB1804800.}}
	{School of Cyber Science and Technology\\
	Beihang University\\
	Beijing, China, 100083}
 {Alfred Hero \sthanks{Supported by ARO grant W911NF-15-1-0479.}}
	{Department of EECS\\
	University of Michigan, Ann Arbor\\
	MI, USA, 48109}

\begin{document}
\maketitle

\begin{abstract}
We establish fundamental limits of tracking a moving target over the unit cube under the framework of 20 questions with measurement-dependent noise. In this problem, there is an oracle who knows the instantaneous location of a target. Our task is to query the oracle as few times as possible to accurately estimate the trajectory of the moving target, whose initial location and velocity is \emph{unknown}. We study the case where the oracle's answer to each query is corrupted by random noise with query-dependent discrete distribution. In our formulation, the performance criterion is the resolution, which is defined as the maximal absolute value between the true location and estimated location at each discrete time during the searching process. We are interested in the minimal resolution of any non-adaptive searching procedure with a finite number of queries and derive approximations to this optimal resolution via the second-order asymptotic analysis.
\end{abstract}
\begin{keywords}
Measurement-dependent noise, resolution, second-order asymptotics, estimation, non-adaptive
\end{keywords}

\section{Introduction}

In the once popular game of ``20 questions''~\cite{wiki20question}, there are two players: one is called oracle and the other is called questioner. The game starts with the oracle who chooses a subject as a secret. The subject chosen for the secret can be arbitrary, e.g., a specific fruit, a certain animal or a particular make of car. The task of the questioner is to figure out the secret of the oracle using at most 20 questions with simple ``yes'' or ``no'' answers. Motivated by this simple game, mathematical models were developed to find optimal query procedures when the oracle can lie. In these models, the secret is replaced by a target random variable which take values in a set, either discrete or continuous. Ulam~\cite{ulam1991adventures} considered the case where the oracle could lie to a finite number of queries while R\'enyi~\cite{renyi1961problem} studied the case where the oracle could lie to each query with certain probability. This mathematical model was later called the Ulam-R\'enyi game and known as the ``20 questions estimation'' problem.

In 20 questions estimation, a query procedure can be either adaptive or non-adaptive. If each query is designed as a function of previous queries and responses to these queries, the query procedure is called adaptive. Otherwise, the query procedure is called non-adaptive. A non-adaptive query procedure has the advantage of lower computation cost and faster execution time over its adaptive counterpart~\cite{jedynak2012twenty}.

Under the goal of minimizing the entropy of the posterior distribution of a target random variable taking values in the unit interval, Jedynak \emph{et al.}~\cite{jedynak2012twenty} proposed both non-adaptive and adaptive procedures which achieve the optimal performance. The non-adaptive procedure in \cite{jedynak2012twenty} is known as the dyadic policy and the adaptive procedure is inspired by the communication strategy with feedback introduced by Horstein~\cite{horstein1963sequential}. Subsequently, the result in \cite{jedynak2012twenty} was generalized to a collaborative case~\cite{tsiligkaridis2014collaborative}, a decentralized case~\cite{tsiligkaridis2015decentralized} and a multi target case~\cite{rajan2015bayesian}. It was realized in later works that other accuracy measures, such as the square loss~\cite{chung2018unequal} and the absolute difference (also known as the resolution~\cite{kaspi2018searching,chiu2016sequential,zhou2019twentyq}) between the estimated and real values of the target, are often better measures of the accuracy. This is because these two measures directly evaluate the performance of a procedure in terms of locating the target. In contrast, if one aims to minimize the posterior entropy of the target variable, then two queries are considered equally important if they can reduce the posterior entropy of the target variable by the same amount. However, it could be true that one query queries the most significant bit while the other queries a much less significant bit in the binary expansion of the target variable. Evidently, the former query is more favorable since it reduces the estimation error more significantly. 

In this paper, we use the 20 questions estimation framework to study the problem of searching for a moving target with unknown initial location and velocity over the unit cube~(cf. \cite[Theorem 3]{kaspi2018searching} for the one-dimensional case). In particular, we focus on the non-adaptive searching procedures and use the resolution (i.e., the maximal absolute difference between the estimated and true values of the initial location or the velocity) as the performance criterion. Our main result is a second-order asymptotic approximation to the performance of optimal non-adaptive searching procedures with finitely many queries. A direct application of our result is searching for a moving target using a sensor network, similar to \cite{tsiligkaridis2014collaborative}. Other applications of 20 questions estimation framework include fault-tolerant communications~\cite{renyi1961problem}, human-in-the-loop decision-making~\cite{tsiligkaridis2014collaborative} and object localization in images~\cite{jedynak2012twenty,rajan2015bayesian}.
\subsection{Main Contributions}

Under mild conditions on the maximal velocity of the target, we derive the second-order asymptotic approximation to the performance of optimal non-adaptive searching procedures with finitely many queries. Intuitively, we derive the accuracy one can achieve with optimal non-adaptive searching procedures as a function of the number of queries and the tolerable excess-resolution (error) probability. We show an interesting phase transition phenomenon, which exhibits a sharp transition on the excess-resolution probability as a function of the decay rate of the target resolution. 

\subsection{Comparison with Most Related Works}
The most closely related work to ours is \cite[Theorem 3]{kaspi2018searching} (see also \cite{6970834}) and \cite{zhou2019twentyq}. In \cite[Theorem 3]{kaspi2018searching}, the authors proposed to study the problem of tracking a moving target with unknown initial location and velocity over the unit circle. The results in \cite[Theorem 3]{kaspi2018searching} are tight when the maximal  velocity tends to zero and thus establishes the asymptotically optimal decay rate for this case. In this paper, we refine \cite[Theorem 3]{kaspi2018searching} by deriving the second-order asymptotic result which provides approximations to the finite blocklength performance. Note that the second-order asymptotic behavior is important since in practical searching problems, one usually tracks a moving target for a limited amount of time, e.g., days or weeks. Our work in this paper is a generalization of the results in \cite{zhou2019twentyq} where one searches for a stationary target. Although both this work and \cite{zhou2019twentyq} use ideas from finite blocklength information theory, our formulation differs significantly from \cite{zhou2019twentyq} because we consider a moving target, which makes the analysis complicated since the location of the target varies over time, introducing combinatorial complexity of the set of all possible trajectories.

\section{Problem Formulation}
\label{sec:pf}
\subsection*{Notation}
Random variables and their realizations are denoted by upper case variables (e.g.,  $X$) and lower case variables (e.g.,  $x$), respectively. All sets are denoted in calligraphic font (e.g.,  $\mathcal{X}$). Let $X^n:=(X_1,\ldots,X_n)$ be a random vector of length $n$. We use $\Phi^{-1}(\cdot)$ to denote the inverse of the cumulative distribution function (cdf) of the standard Gaussian. We use $\bbR$, $\bbR_+$ and $\bbN$ to denote the sets of real numbers, positive real numbers and integers respectively. Given any two integers $(m,n)\in\bbN^2$, we use $[m:n]$ to denote the set of integers $\{m,m+1,\ldots,n\}$ and use $[m]$ to denote $[1:m]$. Given any $(m,n)\in\bbN^2$, for any $m$ by $n$ matrix $\ba=\{a_{i,j}\}_{i\in[m],j\in[n]}$, the infinity norm is defined as $\|\ba\|_{\infty}:=\max_{i\in[m],j\in[n]}|a_{i,j}|$. We use $\calF(\calS)$ to denote the set of all probability density functions on a set $\calS$. All logarithms are base $e$ unless otherwise noted. Finally, we use $\bbo()$ to denote the indicator function.

\subsection{System Model: 20 Questions Estimation Framework}
We state the model for the problem of tracking a moving target with unknown initial location and velocities over a unit cube in the framework of 20 questions estimation. Consider any integer $d\in\bbN$. The initial location of the target is modeled by a $d$-dimensional vector of continuous random variables $\bS=(S_1,\ldots,S_d)$, which take values in the unit cube $[0,1]^d$. Furthermore, it is assumed that the target moves with unknown velocities $\bV=(V_1,\ldots,V_d)$, where for each dimension $i\in[d]$, the velocity takes values in the interval $\calV:=[-v_+,v_+]$. Define the joint probability density function $f_{\bS\bV}$ for the initial location $\bS$ and velocity $\bV$.

Given any initial location $\bs=(s_1,\ldots,s_d)\in[0,1]^d$ and any velocities $\bv=(v_1,\ldots,v_d)\in\calV^d$, at any time $t\in\bbR_+$, for each dimension $i\in[d]$, the  location at the $i$-th dimension of the target $\rml(s_i,v_i,t)$ is the following function of $s_i+tv_i$
\begin{itemize}
\item if $\mathrm{mod}(s_i+tv_i,1)=0$
\begin{align}
\rml(s_i,v_i,t)=1\label{torusc1},
\end{align}
\item if $s_i+tv_i\in\bigcup_{h\in\bbN}[2h,2h+1)$
\begin{align}
\rml(s_i,v_i,t)=s_i+tv_i-\lfloor s_i+tv_i\rfloor
\end{align}
\item otherwise,
\begin{align}
\rml(s_i,v_i,t)=\lceil s_i+tv_i\rceil-(s_i+tv_i)\label{torusc3},
\end{align}
\end{itemize}
where $\mathrm{mod}(a,b)$ refers to the modulo operator and outputs the remainder of $a$ dividing $b$. Note that in the definition of $\rml(s_i,v_i,t)$, we take the ceiling and flooring operations to account for the boundary. For the special case of $d=1$, such a setting is similar to requiring the target to move over the unit circle~\cite[Section V]{kaspi2018searching}. Furthermore, we use $\rml(\bs,\bv,t)$ to denote the $d$-dimensional vector $(\rml(s_1,v_1,t),\ldots,\rml(s_d,v_d,t))$.

It is assumed that there is an oracle who always knows the instantaneous locations of the target. Our task is to pose $n$ queries $\calA^n=(\calA_1,\ldots,\calA_n)\in([0,1]^d)^n$ to the oracle and use the noisy response from the oracle to accurately estimate the initial location and the  velocities of the target. At each time $t\in[n]$, we query the oracle whether the target lies in the region $\calA_t\in[0,1]^d$. The correct answer $X_t=\bbo(\rml(\bS,\bV,t)\in\calA_t)$ is corrupted by a measurement-dependent noise as $X_t$ is transmitted through a channel with transition matrix $P_{Y|X}^{\calA_i}$ (see Section \ref{sec:mdc} for definitions) to yield a noisy response $Y_i$ which takes value in the alphabet $\calY$. Given the noisy responses to all the $n$ queries, we use a decoder $g:\calY^n\to[0,1]^d\times\calV^d$ to estimate the initial location $\bS$ and velocities $\bV$.

We consider non-adaptive searching procedures where each query $\calA_t$ is designed independently of each other. In adaptive searching procedure~\cite{chiu2019noisy}, at each time $t\in[n]$, the design of query $\calA_t$ depends on all previous queries and noisy responses to these queries, i.e., $\{\calA_j,Y_j\}_{j\in[t-1]}$. 

\vspace{-0.5em}
\subsection{The Measurement-Dependent Channel}
\label{sec:mdc}
\vspace{-0.5em}
In this subsection, we briefly describe the measurement-dependent channel~\cite{kaspi2018searching,chiu2016sequential}, also known as a channel with state~\cite[Chapter 7]{el2011network}. Given a sequence of queries $\calA^n\subseteq([0,1]^d)^n$, the channel from the oracle to the player is a memoryless channel whose transition probabilities are functions of the queries. Specifically, for any $(x^n,y^n)\in\{0,1\}^n\times\calY^n$,
\begin{align}
P_{Y^n|X^n}^{\calA^n}(y^n|x^n)
&=\prod_{i\in[n]}P_{Y|X}^{\calA_i}(y_i|x_i),
\end{align}
where $P_{Y|X}^{\calA_i}$ denotes the transition probability of the channel which depends on the $i$-th query $\calA_i$. Given any query $\calA\subseteq[0,1]^d$, define the size $|\calA|$ of $\calA$ as its Lebesgue measure, i.e., $|\calA|=\int_{t\in\calA}\rmd t$. Throughout the paper, we assume that the measurement-dependent channel $P_{Y|X}^{\calA}$ depends on the query $\calA$ only through its size, i.e., $P_{Y|X}^{\calA}$ is equivalent to a channel with state $P_{Y|X}^q$ where the state $q$ is a function of $|\calA|$. Throughout the paper, we assume that $q=f(|\calA|)$ where the function $f:[0,1]\to\bbR_+$ is a bounded Lipschitz continuous function with parameter $K$, i.e., $|f(q_1)-f(q_2)|\leq K|q_1-q_2|$ and $\max_{q\in[0,1]}f(q)<\infty$.

For any $q\in[0,1]$, any $\xi\in(0,\min(q,1-q))$, we assume the measurement-dependent channel is continuous in the sense that there exists a constant $c(q)$ depending on $q$ only such that
\begin{align}
\max\left\{\left\|\log\frac{P_{Y|X}^q}{P_{Y|X}^{{q+\xi}}}\right\|_{\infty},\left\|\log\frac{P_{Y|X}^q}{P_{Y|X}^{q-\xi}}\right\|_{\infty}\right\}\leq c(q)\xi\label{assump:continuouschannel}.
\end{align}

An example is given as follows.
\begin{definition}
\label{def:mdBSC}
Given any $\calA\subseteq[0,1]$, a channel $P_{Y|X}^{\calA}$ is said to be a measurement-dependent Binary Symmetric Channel (BSC) with parameter $\zeta\in(0,1]$ if $\calX=\calY=\{-1,1\}$ and 
\begin{align}
P_{Y|X}^{\calA}(y|x)=(\zeta f(|\calA|))^{\bbo(y\neq x)}(1-\zeta f(|\calA|))^{\bbo(y=x)}.
\end{align}
\end{definition}
This definition generalizes \cite[Theorem 1]{kaspi2018searching}, where the authors considered a measurement-dependent BSC with parameter $\zeta=1$ and the function $f(|\calA|)=|\calA|$. Note that the output of a measurement dependent BSC with parameter $\zeta$ is the same as the input with probability $1-\zeta f(|\calA|)$ and flipped with probability $\zeta f(|\calA|)$. It can be verified that the constraint in \eqref{assump:continuouschannel} is satisfied for the measurement-dependent BSC.

\subsection{Definition of the Fundamental Limit}
Recall that we consider tracking a target with unknown initial location and velocity over a unit cube of dimension $d\in\bbN$. A non-adaptive searching procedure is defined as follows.
\begin{definition}
\label{def:procedure:moving}
Given any $n\in\bbN$, $\delta\in\bbR_+$ and $\varepsilon\in[0,1]$, an $(n,d,\delta,\varepsilon)$-non-adaptive searching procedure consists of 
\begin{itemize}
\item $n$ queries $\calA^n$, where for each $i\in[n]$, we query whether the target locates in a Lebesgue measurable subset $\calA_i\in[0,1]^d$
\item and an estimator $g:\calY^n\to[0,1]^d\times\calV^d$
\end{itemize}
such that the excess-resolution probability satisfies
\begin{align}
\rmP_\rme(n,d,\delta)
\nn&:=\sup_{f_{\bS\bV}}\Pr\Big\{\exists~i\in[d]:\\*
&\max_{t\in[0:n]}|(\hatS_i+t\hatV_i)-(S_i+tV_i)|>\delta\Big\}\leq \varepsilon\label{evaluate},
\end{align}
where $\hatS_i$ is the estimate for the $i$-th dimension $S_i$ of the initial location $\bS$ and $\hatV_i$ is the estimate for the velocity $V_i$ in the $i$-th dimension of the velocity $\bV$. Furthermore,  $(\hatS_1,\ldots,\hatS_d,\hatV_1,\ldots,\hatV_d)$ is generated by the decoder $g(Y^n)$.
\end{definition}
Note that for any $(n,d,\delta,\varepsilon)$ searching procedure, with probability of at least $1-\varepsilon$, regardless of the joint distributions of the initial location and velocity, we can accurately estimate the trajectories of the target within resolution $\delta$ in each dimension any any searching time during the whole process. When specialized to the case of $d=1$, our formulation in \eqref{evaluate} imposes a stronger requirement on the tracking performance, compared with \cite{kaspi2018searching}, where the estimates $\hatS$ and $\hatV$ are both constrained to be within $\delta$ of the actual values $S$ and $V$. As a result, in the formulation of \cite{kaspi2018searching}, the maximal difference between the estimated and true locations of the target can be as large as $n\delta$, where $n$ is the number of queries.

We are interested in the performance of optimal searching procedures that are evaluated by the minimal achievable resolution defined as
\begin{align}
\delta^*(n,d,\varepsilon)
\nn&:=\inf\{\delta\in\bbR_+:\exists~\mathrm{an~}(n,d,\delta,\varepsilon)\\*
&\mathrm{-non}\mathrm{-adaptive}\mathrm{~tracking}\mathrm{~procedure}\}.
\end{align}
A dual definition is the sample complexity, which is the minimal number of queries to achieve a given resolution. However, it suffices to study the minimal resolution $\delta^*(n,d,\varepsilon)$ since the sample complexity is in fact a function of $\delta^*(n,d,\varepsilon)$.

\section{Main Results and Discussions}
\label{sec:results}

\subsection{Preliminaries}
\label{sec:preliminaries}
We present necessary preliminary definitions.  Given any $(p,q)\in[0,1]^2$, let $P_Y^{p,q}$ be the marginal distribution on $\calY$ induced by the Bernoulli distribution $P_X=\mathrm{Bern}(p)$ and the measurement-dependent channel $P_{Y|X}^{q}$. Furthermore, define the following information density
\begin{align}
\imath_{p,q}(x;y)&:=\log\frac{P_{Y|X}^q(y|x)}{P_Y^{p,q}(y)},~\forall~(x,y)\in\calX\times\calY.
\end{align}
Correspondingly, for any $(x^n,y^n)\in\calX^n\times\calY^n$, we define
\begin{align}
\imath_{p,q}(x^n;y^n)
&:=\sum_{t\in[n]}\imath_{p,q}(x_t;y_t)\label{def:ix^ny^n}
\end{align}
as the empirical mutual information between $x^n$ and $y^n$.

\subsection{Second-Order Asymptotics}
In this subsection, we derive the second-order approximation to the fundamental limit $\delta^*(n,d,\varepsilon)$. We consider maximal velocity $v_+$, which ensures that the total moving distance is sub-linear in the number of queries, i.e., $nv_+=o(n)$. We show that exact asymptotic characterization can be obtained under such an assumption and furthermore, exact second-order approximation can be obtained if the moving speed $v_+$ is further constrained such that $nv_+=o(\sqrt{n})$.

To present our results, we need the following definitions. For any measurement-dependent channel $\{P_{Y|X}^q\}_{q\in[0,1]}$, let
\begin{align}
C:=\max_{p\in(0,1)} \mathbb{E}[\imath_{p,f(p)}(X;Y)]\label{def:c},
\end{align}
where $(X,Y)\sim \mathrm{Bern}(p)\times P_{Y|X}^{f(p)}$. Let $\calP_{\rm{ca}}$ be the set of values $p$ achieving $C$. Furthermore, given any $p\in(0,1)$, let
\begin{align}
\rmV_p&:=\mathrm{Var}[\imath_{p,f(p)}(X;Y)]\label{def:vp},\\
\rmT_p&:=\mathbb{E}[|\imath_{p,f(p)}(X;Y)-\mathbb{E}[\imath_{p,f(p)}(X;Y)]|^3]\label{def:tp}.
\end{align}
Finally, given any $\varepsilon\in(0,1)$, let
\begin{align}
\rmV_{\varepsilon}
&=\left\{
\begin{array}{ll}
\max_{p\in\calP_{\rm{ca}}}\rmV_p&\mathrm{if~}\varepsilon\leq 0.5,\\
\min_{p\in\calP_{\rm{ca}}}\rmV_p&\mathrm{otherwise}.
\end{array}
\right.\label{def:v}
\end{align}

Note that since we consider channels with finite output alphabet, for any $p\in\calP_{\rm{ca}}$, the third absolute moment of $\imath_{p,f(p)}(X;Y)$ is finite~\cite[Lemma 47]{polyanskiy2010finite}.
\begin{figure}[htbp]
\centering
\includegraphics[width=.9\columnwidth]{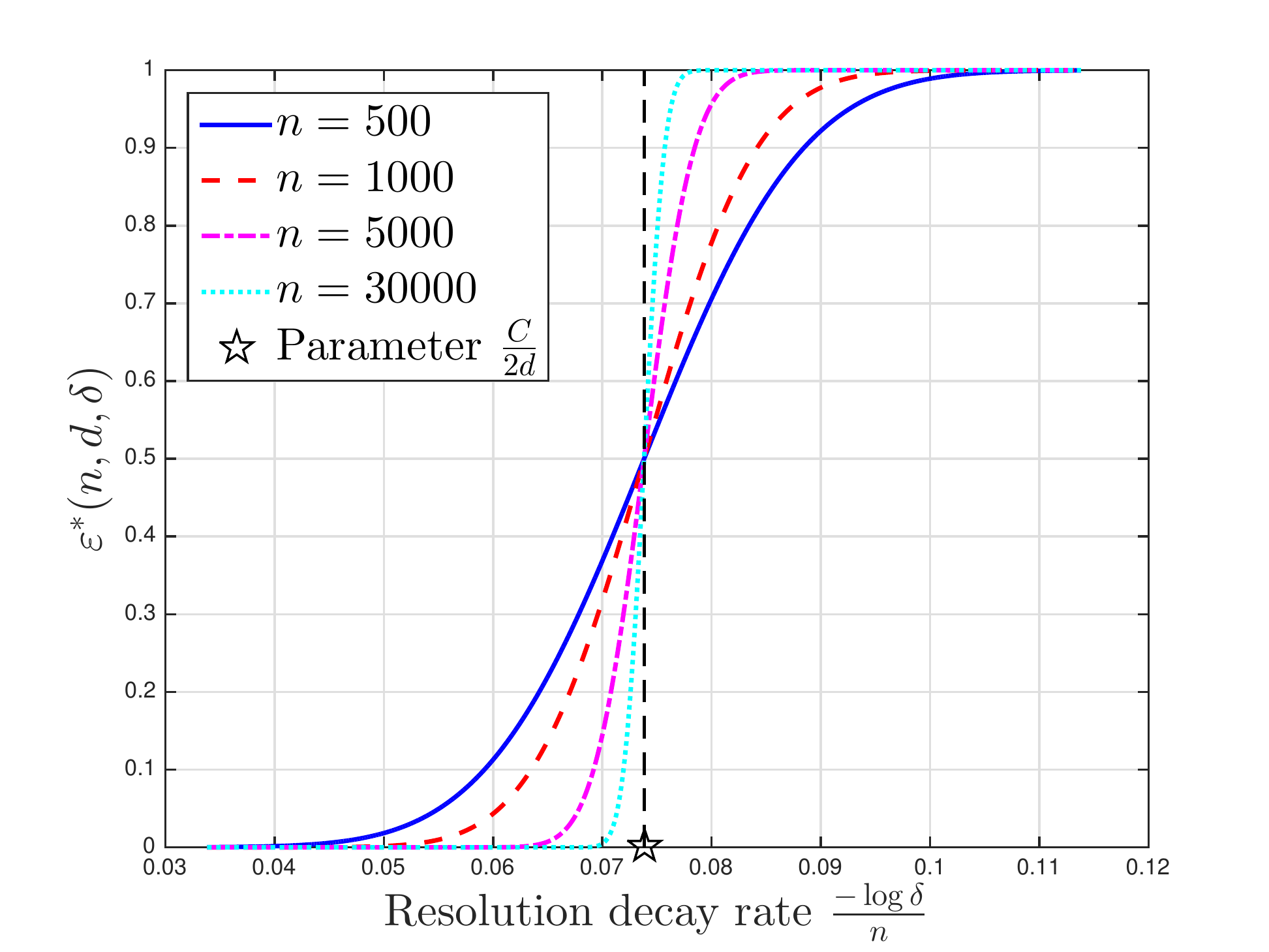}
\caption{Illustration of the phase transition phenomenon of optimal non-adaptive searching procedures for a measurement-dependent BSC with $\zeta=0.2$. The Lipschitz continuous function is  $f(q)=2q+0.5$. The pentagram star denotes the critical phase transition threshold.}
\label{illus:phasetransition}
\end{figure}
\begin{theorem}
\label{mainresult}
For any $\varepsilon\in(0,1)$ and finite $d\in\bbN$, the minimal achievable resolution $\delta^*(n,d,\varepsilon)$ satisfies that
\begin{itemize}
\item if $nv_+=O(n^t)$ for $t\in[0.5,1)$,
\begin{align}
-2d\log\delta^*(n,d,\varepsilon)
=nC+O(nv_+);
\end{align}
\item if $nv_+=O(n^t)$ for $t\in[0,0.5)$
\begin{align}
-2d\log\delta^*&(n,d,\varepsilon)
\nn=nC+\sqrt{n\rmV_{\varepsilon}}\Phi^{-1}(\varepsilon)\\*
&\qquad+O(\max\{nv_+,\log n\});
\end{align}

\end{itemize}
\end{theorem}
The proof and further discussions of Theorem \ref{mainresult} are provided in our extended version~\cite{zhou2020tracking}. Theorem \ref{mainresult} refined the asymptotic result in \cite[Theorem 3]{kaspi2018searching}. For maximal velocity satisfying certain \emph{practical} conditions, we derived an approximation to the achievable resolution of optimal non-adaptive searching procedures that employ a finite number of queries. 

Furthermore, Theorem \ref{mainresult} implies the phase transition phenomenon governing optimal non-adaptive searching procedures. Specifically, if we let $\varepsilon^*(n,d,\delta)$ be the minimal excess-resolution probability of any non-adaptive query procedure, when the maximal velocity $v_+$ is such that $nv_+=o(\sqrt{n})$, our result in Theorem \ref{mainresult} implies that for $n$ sufficiently large,
\begin{align}
\varepsilon^*(n,d,\delta)=\Phi\left(\frac{-d\log\delta-nC}{\sqrt{nV_\varepsilon}}\right)+o(1)\label{phaset}.
\end{align}
Note that \eqref{phaset} implies the existence of a phase transition phenomenon for the minimal excess-resolution probability as a function of the resolution decay rate $-\frac{\log \delta}{n}$. In particular,  when the target resolution decay rate $\frac{-\log\delta}{n}$ is strictly greater than $\frac{C}{2d}$, the minimal excess-resolution probability tends to \emph{one} as the number of queries $n$ tends to infinity. On the other hand, when the target resolution decay rate is strictly less than the critical rate $\frac{C}{2d}$, the excess-resolution probability \emph{vanishes} as the number of queries $n$ increases. We numerically illustrate the phase transition phenomenon for a measurement-dependent BSC in Figure \ref{illus:phasetransition} for the case of $d=1$.

\section{Conclusion}
\vspace{-0.5em}
We derived a second-order asymptotic approximation to the performance of optimal non-adaptive searching for a moving target with unknown initial location and velocities using finitely many queries. In future, one can propose low complexity searching procedures and compare their empirical performances with our derived benchmarks. It is also interesting to relax the ``torus'' constraint in Eq. \eqref{torusc1} to \eqref{torusc3} on the moving target and study the fundamental limit of optimal searching strategies. Finally, one can study adaptive searching procedures and analyze the benefit of adaptivity.

\newpage
\bibliographystyle{IEEEbib}
\bibliography{IEEEfull_lin}

\begin{thebibliography}{10}

\bibitem{wiki20question}
Wikipedia,
\newblock ``Twenty questions,'' 05 2020.

\bibitem{ulam1991adventures}
Stanislaw~M Ulam,
\newblock {\em Adventures of a Mathematician},
\newblock Univ. of California Press, 1991.

\bibitem{renyi1961problem}
Alfr{\'e}d R{\'e}nyi,
\newblock ``On a problem of information theory,''
\newblock {\em MTA Mat. Kut. Int. Kozl. B}, vol. 6, pp. 505--516, 1961.

\bibitem{jedynak2012twenty}
Bruno Jedynak, Peter~I Frazier, and Raphael Sznitman,
\newblock ``Twenty questions with noise: Bayes optimal policies for entropy
  loss,''
\newblock {\em J. Appl. Probab.}, vol. 49, no. 1, pp. 114--136, 2012.

\bibitem{horstein1963sequential}
Michael Horstein,
\newblock ``Sequential transmission using noiseless feedback,''
\newblock {\em IEEE Trans. Inf. Theory}, vol. 9, no. 3, pp. 136--143, 1963.

\bibitem{tsiligkaridis2014collaborative}
Theodoros Tsiligkaridis, Brian~M Sadler, and Alfred~O. Hero,
\newblock ``Collaborative 20 questions for target localization,''
\newblock {\em IEEE Trans. Inf. Theory}, vol. 60, no. 4, pp. 2233--2252, 2014.

\bibitem{tsiligkaridis2015decentralized}
Theodoros Tsiligkaridis, Brian~M Sadler, and Alfred~O. Hero,
\newblock ``On decentralized estimation with active queries,''
\newblock {\em IEEE Trans. Signal Process.}, vol. 63, no. 10, pp. 2610--2622,
  2015.

\bibitem{rajan2015bayesian}
Purnima Rajan, Weidong Han, Raphael Sznitman, Peter Frazier, and Bruno Jedynak,
\newblock ``Bayesian multiple target localization,''
\newblock {\em J. Mach. Learn. Res.}, vol. 37, pp. 1945--1953, 2015.

\bibitem{chung2018unequal}
Hye~Won Chung, Brian~M Sadler, Lizhong Zheng, and Alfred~O. Hero,
\newblock ``Unequal error protection querying policies for the noisy 20
  questions problem,''
\newblock {\em IEEE Trans. Inf. Theory}, vol. 64, no. 2, pp. 1105--1131, 2018.

\bibitem{kaspi2018searching}
Yonatan Kaspi, Ofer Shayevitz, and Tara Javidi,
\newblock ``Searching with measurement dependent noise,''
\newblock {\em IEEE Trans. Inf. Theory}, vol. 64, no. 4, pp. 2690--2705, 2018.

\bibitem{chiu2016sequential}
Sung-En Chiu and Tara Javidi,
\newblock ``Sequential measurement-dependent noisy search,''
\newblock in {\em IEEE ITW}, 2016, pp. 221--225.

\bibitem{zhou2019twentyq}
Lin Zhou and Alfred Hero,
\newblock ``Resolution limits of noisy non-adaptive 20 questions estimation,''
\newblock {\em To appear in IEEE Trans. Inf. Theory}, 2021.

\bibitem{6970834}
Y.~{Kaspi}, O.~{Shayevitz}, and T.~{Javidi},
\newblock ``Searching with measurement dependent noise,''
\newblock in {\em IEEE ITW}, Nov 2014, pp. 267--271.

\bibitem{chiu2019noisy}
Sung-En Chiu,
\newblock {\em Noisy Binary Search: Practical Algorithms and Applications},
\newblock Ph.D. thesis, UC San Diego, 2019.

\bibitem{el2011network}
Abbas El~Gamal and Young-Han Kim,
\newblock {\em Network Information Theory},
\newblock Cambridge University Press, 2011.

\bibitem{polyanskiy2010finite}
Yury Polyanskiy, H~Vincent Poor, and Sergio Verd\'u,
\newblock ``Channel coding rate in the finite blocklength regime,''
\newblock {\em IEEE Trans. Inf. Theory}, vol. 56, no. 5, pp. 2307--2359, 2010.

\bibitem{zhou2020tracking}
Lin Zhou and Alfred Hero,
\newblock ``Searching for a moving target with unknown initial location and
  velocity,''
\newblock Tech. {R}ep., https://zl19920612.wixsite.com/linzhou/publications,
  2020.

\end{thebibliography}
\end{document}